# Phase-Sensitive Nonlinear X-Ray Response in a Charge-Density-Wave Material

S. Sofer[1], G. J. Man[1*], A. Bombardi[4], and S. Shwartz[1]

[1]*Physics Department and Institute of Nanotechnology, Bar-Ilan University, Ramat Gan, 52900 Israel*

[4] *Diamond Light Source, Harwell Science and Innovation Campus, Didcot OX11 0DE, United Kingdom*

*\* Currently at Man Labs Technologies LLC, Dallas, Texas 75219, USA*

We report a phase-sensitive nonlinear x-ray response in the charge-density-wave material 1T-TaS$_2$, revealed through x-ray parametric down-conversion into the ultraviolet. Extending nonlinear x-ray wave mixing beyond conventional crystalline systems to a correlated quantum material, we employ reciprocal-lattice phase matching to isolate distinct Fourier components of the nonlinear susceptibility. By selecting a fundamental reciprocal-lattice vector and a stacking-sensitive half-integer reciprocal-lattice vector, we probe the response across the nearly commensurate and incommensurate charge-density-wave phases. Tuning the ultraviolet photon energy through Ta O-shell resonances uncovers pronounced Fourier-component and phase-dependent resonant structure, indicating that the stacking-related nonlinear susceptibility couples differently to Ta-centered resonant states than the average lattice response. Remarkably, the nonlinear signal is strongly enhanced in the nearly commensurate phase despite weaker Bragg diffraction, demonstrating that the nonlinear susceptibility provides information inaccessible to linear probes. These results establish nonlinear x-ray spectroscopy as a phase-sensitive and orbital-selective probe of electronic reconstruction in quantum materials.

Nonlinear x-ray optics extends the concepts of nonlinear optical spectroscopy to the atomic length scales and core-level energies accessible with x rays. In crystals, reciprocal-lattice vectors provide momentum selectivity that enables direct access to specific Fourier components of the electronic response. At the same time, wave-mixing processes involving long wavelength fields (typically in the UV or optical frequencies) directly probe low-energy electronic degrees of freedom [1-12]. Together, these capabilities offer a route to momentum-resolved nonlinear spectroscopy that can separate lattice, stacking, and electronic contributions to the response.

Despite this promise, nonlinear x-ray wave-mixing experiments have so far been demonstrated primarily in simple, weakly correlated materials [1-11] In such systems, the nonlinear response closely follows the average lattice structure, and the experiments rely on high crystal quality to separate the weak nonlinear signal from the much stronger Bragg diffraction. As a result, applying nonlinear x-ray techniques to quantum materials, where reduced coherence, structural complexity, and intertwined electronic orders are intrinsic, has remained a significant experimental challenge.

A central nonlinear x-ray process in this context is x-ray parametric down-conversion (PDC), which generates a correlated photon pair consisting of an x-ray signal photon and a longer-wavelength idler photon through a phase-matched wave-mixing process. In a crystal, momentum conservation is mediated by a reciprocal-lattice vector $\vec{G}$, enabling selective access to specific Fourier components of the nonlinear susceptibility. By tuning the idler photon energy across ultraviolet/optical resonances via phase matching and signal-energy filtering, PDC provides a practical route to resonant nonlinear spectroscopy without changing the pump energy. Although the PDC signal is intrinsically weak and lies close to the much stronger elastic background, a recent work has demonstrated its observability even in reduced-coherence systems [12], opening the possibility of applying nonlinear x-ray wave mixing to correlated quantum materials.

The layered compound 1T-TaS$_2$ hosts multiple charge-density-wave (CDW) phases and a long-standing debate over the microscopic origin of its insulating state [13-16]. Upon cooling, 1T-TaS$_2$ undergoes transitions from an incommensurate CDW (ICCDW) phase to a nearly commensurate CDW (NCCDW) phase, and finally to a commensurate CDW (CCDW) [15], accompanied by changes in stacking and electronic structure. Interlayer dimerization associated with the ordered star of David (SoD) array produces characteristic half-integer reflections along the out-of-plane direction, while the disappearance of these reflections is commonly interpreted as a collapse of the dimerized stacking [13,15,17,18]. Whether this stacking order or correlation-driven Mott localization is primarily responsible for the insulating gap remains actively debated [16,19,20]. Because stacking and average lattice contributions are encoded in different reciprocal-lattice components, selectively probing Fourier components associated with fundamental and stacking-sensitive vectors provides a direct means to disentangle their respective contributions to the electronic reconstruction across the CDW transitions.

In this letter we demonstrate x-ray PDC into UV in 1T-TaS$_2$ and show that the nonlinear response is directly sensitive to both the CDW phase and the stacking order. By comparing PDC signals associated with a fundamental reciprocal lattice vector $\vec{G} = (0,0,4)$ and a stacking-sensitive half-integer vector $\vec{G}=(0,0,7/2)$, we isolate distinct Fourier components of the nonlinear electronic susceptibility. Tuning the idler photon energy through the Ta O-shell resonances reveals pronounced, Fourier-component and phase-dependent resonant structure. Of importance, the PDC intensity exhibits trends opposite to those of elastic scattering across the NCCDW–ICCDW transition, identifying the nonlinear susceptibility as an independent probe of the electronic reconstruction. These results establish x-ray PDC as a phase-sensitive and orbital-selective probe of electronic phenomena in quantum materials.

All measurements were carried out at beamline I16 of Diamond Light Source [21]. A schematic of the setup and the PDC geometry is shown in Fig. 1. The incident x-ray beam was monochromatic and collimated, defined by a Si(111) monochromator, and set to a pump energy of $E_p = 9$ keV for the measurements reported here. The nonlinear medium was a <0,0,1> oriented single-crystal 1T-TaS$_2$ sample manufactured by HQ Graphene, mounted on the I16 diffractometer inside a cryostat on a rotating stage. This geometry provided access to the (0,0,4) Bragg reflection and, when present, the stacking-sensitive half-integer (0,0,7/2) reflection, corresponding to reciprocal-lattice vectors $\vec{G}$ used for phase matching. Data were collected at stabilized temperatures of 100 K, 200 K, and 400 K, all reached on cooling.

To isolate the weak down-converted x-ray signal photons from the much stronger elastically scattered background, a three-bounce Si(111) crystal analyzer was used as a narrow-band energy filter, and the transmitted photons were recorded by a two-dimensional Merlin detector. For each selected idler photon energy, the PDC intensity was obtained from the crystal rocking curves around the phase-matching condition, with the peak intensity taken as the signal. Phase matching was verified by comparing the measured peak positions with calculated phase-matching angles obtained from the momentum conservation condition $\vec{k}_p + \vec{G} = \vec{k}_s + \vec{k}_{id}$, where $\vec{k}_p, \vec{k}_s, \vec{k}_{id}$ are the wave vectors of the pump, signal and idler, respectively. Typical PDC count rates at the phase-matching condition were on the order of $10^2$–$10^5$ counts s$^{-1}$, compared with the elastic scattering peak intensities of $10^7$–$10^9$ counts s$^{-1}$, indicating that the measurement sensitivity is limited primarily by residual elastic background.

We first characterized the phase-matching conditions by measuring PDC rocking curves associated with the $\vec{G}$=(0,0,4) and stacking-sensitive $\vec{G}$=(0,0,7/2) in different CDW phases of 1T-TaS$_2$. For each selected idler energy, the analyzer was set to the corresponding signal energy, and the crystal was rocked through the phase-matching condition. The PDC peak angle was calculated for different regions of interest (ROI) on the detector, corresponding to different detector angles. Each ROI had a width of 0.01 degrees, centered at a distinct detector angle.

Figure 2 compares the measured rocking-curve peak angles at an idler energy of 40 eV with the calculated phase-matching angles, showing agreement within the experimental uncertainties, confirming x-ray PDC. Additional rocking curves and measurements at other idler energies are provided in the Supplementary Information.

In the NCCDW phase we observed well-defined rocking curves for both the $\vec{G}$=(0,0,4) and $\vec{G}$=(0,0,7/2). In the ICCDW phase, where the half-integer order collapses, PDC was observed only for $\vec{G}$=(0,0,4). No PDC signal was detected in the CCDW phase within experimental sensitivity. We attribute this to the strongly enhanced elastic scattering in this phase, which increased background and reduced the detectable nonlinear signal.

Next, we show that x-ray PDC in 1T-TaS$_2$ probes electronic reconstruction beyond Bragg diffraction. Because the nonlinear response follows the electronic susceptibility, it develops phase-dependent spectral structure across the CDW transitions [8,9,10,22]. Figure 3 shows the idler-energy dependence of the PDC intensity in the NCCDW phase for the (0,0,4) and (0,0,7/2) vectors and for the (0,0,4) vector in the ICCDW phase, together with the Ta O$_3$, O$_2$, and O$_1$ resonance energies [23].

To interpret the observed PDC spectra, we use a physically motivated model in which the nonlinear signal at each $\vec{G}$ is the coherent sum of slowly varying non-resonant background and resonant contributions from the Ta O-shell transitions. The resonant part is written as Fano-type amplitudes, capturing the asymmetric line shapes observed near core-level resonances [7,24-25]. The full model and the fitting parameters are described in the supplementary information. The model curves are overlaid on the data in Fig. 3.

As shown in Fig. 3, the model accurately reproduces the measured spectra for both fundamental and stacking-sensitive reciprocal-lattice vectors across the NCCDW and ICCDW phases. Pronounced spectral structure is observed near the Ta O-shell resonances, indicating that the PDC process is orbital selective, and can probe individual intermediate states. For the (0,0,4) vector, a prominent feature appears near the Ta $O_3$ resonance in both the NCCDW and ICCDW phases. For the stacking-sensitive (0,0,7/2) vector, we instead observe a weaker but clearly resolved feature near the Ta $O_2$ resonance. Together, the data and fits show that the PDC probes the coherent electronic structure at the selected reciprocal-lattice vectors, with the fitting parameters encoding Fourier component and phase-dependent coupling to the Ta O-shell resonances. The distinct spectral profiles across the three spectra therefore reveal how different Fourier components of the CDW couple to core-level electronic states.

To further isolate electronic information unique to x-ray PDC we analyze intensity ratios, which suppress common experimental factors and highlight genuine changes in the nonlinear response. Figure 4(a) shows the ratio between the PDC intensities associated with the stacking-sensitive $\vec{G}$=(0,0,7/2) and the fundamental $\vec{G}$=(0,0,4) in the NCCDW phase. Figure 4(b) shows the ratio the (0,0,4) PDC intensities in the NCCDW and the ICCDW phase.

For elastic scattering, the intensity ratio between the lattice (0,0,4) and superlattice (0,0,7/2) peaks is $10^{-2} - 10^{-3}$, consistent with our measurements [15]. Far from resonances, we observed a comparable ratio in PDC. However, near the Ta O-edge resonances the (0,0,4) signal is strongly enhanced, whereas the stacking-sensitive (0,0,7/2) signal exhibits a much weaker resonant response. The observed minima in the ratio between the intensities are consistent with destructive interference from the bilayer CDW stacking sequence, suggesting an antiphase relationship between the charge modulations in neighboring bilayers.

The phase comparison is even more striking. The (0,0,4) PDC intensity is enhanced in NCCDW relative to ICCDW, even though the elastic (0,0,4) Bragg intensity is larger in ICCDW. This contrast points to an intrinsic change in the resonant nonlinear susceptibility, rather than a trivial change in diffraction strength. The strong energy dependence further indicates that the ICCDW–NCCDW transition qualitatively modifies the resonant electronic response that drives the PDC process, likely by modifying coupling to low-energy electronic states tied to the CDW order. Together, these ratios show that PDC is phase sensitive and reveals CDW physics beyond elastic scattering.

In summary, we have demonstrated x-ray-to-UV PDC in the quantum material 1T-$TaS_2$, and shown that nonlinear x-ray spectroscopy can isolate symmetry and stacking-selective Fourier components of the CDW response. Spectroscopy through the Ta O-shell resonances reveals a pronounced, phase-dependent enhancement of the nonlinear signal at the fundamental reciprocal-lattice vector (0,0,4), whereas the stacking-sensitive half-integer vector (0,0,7/2) exhibited strongly reduced resonant contrast. Even more strikingly, the nonlinear PDC intensity evolves opposite to the elastic Bragg intensity across the ICCDW-NCCDW transition, identifying the nonlinear susceptibility as an independent probe of electronic structure beyond linear diffraction.

Improved energy resolution (utilizing bent crystal analyzer, for an example) will enable access to finer resonant structure within the electronic bands, further enhancing the orbital and momentum selectivity of the technique. Extending the approach to additional reciprocal-lattice vectors, both conventional and half-integer, offers a direct route to disentangling the microscopic origin of stacking and its role in the CDW landscape. More broadly, the framework established here is readily extendable to other correlated and low-dimensional quantum materials, providing a general momentum-resolved nonlinear spectroscopic route to separating lattice, orbital, and stacking contributions to electronic reconstruction.

This work was supported by the Israel Science Foundation (ISF) (IL), Grant No. 2208/24. We acknowledge Diamond Light Source for the provision of time on Beamline I16 under Proposal 34685. This work was supported in part by the Zuckerman STEM Leadership Program. We thank Efrat Shimshoni, Beena Kalisky, Jonathan Ruhman, Amos Sharoni, and Eran Maniv for helpful discussions.

References

1. I. Freund and B. Levine, Physical Review Letters **26**, 156 (1971).

2. I. Freund, Chemical Physics Letters. **12**, 583-588 (1972).

3. P. Eisenberger and S. McCall, Physical Review A **3**, 1145 (1971).

4. T. Glover, D. Fritz, M. Cammarata, T. Allison, S. Coh, J. Feldkamp, H. Lemke, D. Zhu, Y. Feng, R. Coffee, M. Fuchs, S. Ghimire, J. Chen, S. Shwartz, D. Reis, S. Harris and J. Hastings, Nature **488**, 603-608 (2012).

5. K. Tamasaku and T. Ishikawa, Physical Review Letters **98**, 244801 (2007).

6. K. Tamasaku, K. Sawada and T. Ishikawa, Physical Review Letters **103**, 254801 (2009).

7. K. Tamasaku, K. Sawada, E. Nishibori and T. Ishikawa, Nature Physics **7**, 705-708 (2011).

8. A. Schori, C. Bömer, D. Borodin, S. Collins, B. Detlefs, M. Moretti Sala, S. Yudovich and S. Shwartz, Physical Review Letters **119**, 253902 (2017).

9. D. Borodin, A. Schori, J. Rueff, J. Ablett and S. Shwartz, Physical Review Letters **122**, 023902 (2019).

10. S. Sofer, O. Sefi, E. Strizhevsky, H. Aknin, S. Collins, G. Nisbet, B. Detlefs, C. Sahle and S. Shwartz, Nature Communications 10, 1-8 (2019).

11. C. Ornelas-Skarin, T. Bezriadina, M. Fuchs, S. Ghimire, J. B. Hastings, Q. L. Nguyen, G. de la Peña, T. Sato, S. Shwartz, M. Trigo, D. Zhu, D. Popova-Gorelova and D. A. Reis, Physical Review X 16, 011006 (2026).

12. H. Aknin, O. Sefi, D. Borodin, E. Strizhevsky, J.-P. Rueff, J. M. Ablett and S. Shwartz, Physical Review Research **7**, 043226 (2025).

13. J. A. Wilson, F. J. Di Salvo and S. Mahajan, Physical Review Letters 32, 882 (1974).

14. D. D. Darancet, A. J. Millis and C. A. Marianetti, Physical Review B 90, 045134 (2014).

15. Y. D. Wang, W. L. Yao, Z. M. Xin, T. T. Han, Z. G. Wang, Nature Communications **11**, 4215 (2020)

16. Y. Wang, W. Yao, Z. Xin, T. Han and Z. Wang, Nature Communications **15**, 3425 (2024)

17. A. Spijkerman, J. L. de Boer, A. Meetsma, G. A. Wiegers and S. van Smaalen, Physical Review B **56**, 13757 (1997)

18. T. Ritschel, J. Trinckauf, K. Koepernik, B. Büchner, M. v. Zimmermann, H. Berger, Y. I. Joe, P. Abbamonte and J. Geck, Nature Physics 11, 328 (2015).

19. S.-H. Lee, J. S. Goh and D. Cho, Physical Review Letters **122**, 106404 (2019)

20. C. J. Butler, M. Ligges, D. Golež, S. Peli, A. F. Kemper, U. Bovensiepen, M. Eckstein and M. Grioni, Nature Communications **11**, 2477 (2020).

21. S. P. Collins, A. Bombardi, A. R. Marshall, J. H. Williams, G. Barlow, A. G. Day, M. R. Pearson, R. J. Woolliscroft, R. D. Walton, G. Beutier, and G. Nisbet, AIP Conf. Proc. 1234, 303 (2010).

22. R. Cohen and S. Shwartz, Physical Review Research 1, 033133 (2019).

23. P. Buabthong, N. Becerra Stasiewicz, S. Mitrovic and N. S. Lewis, Surface Science Spectra 24, 024001 (2017).

24. M. F. Limonov, M. V. Rybin, A. N. Poddubny and Y. S. Kivshar, Nature Photonics **11**, 543-554 (2017).

25. K. P. Heeg, C. Ott, D. Schumacher, H.-C. Wille, R. Röhlsberger, T. Pfeifer, and J. Evers, Physical Review Letters **114**, 207401 (2015).

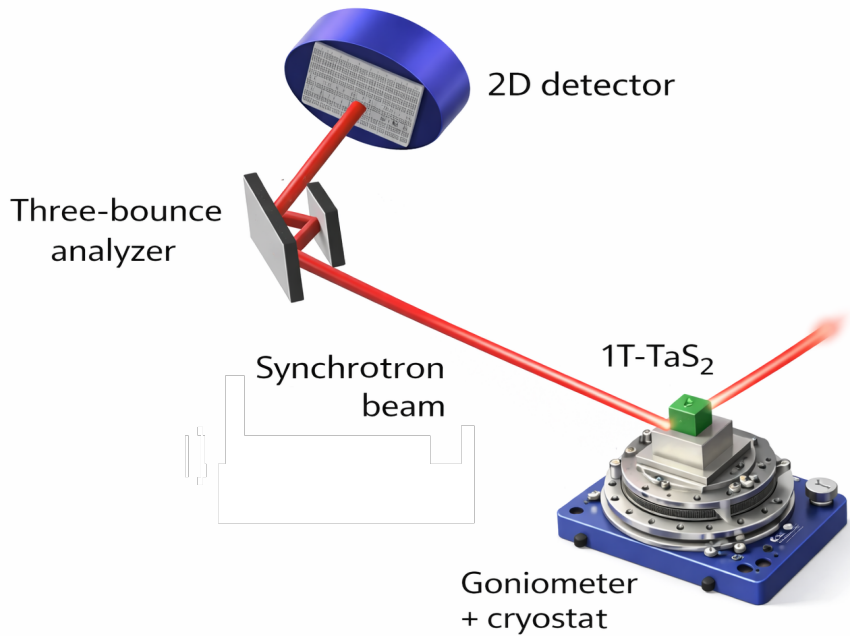

**Fig 1:** Experimental geometry for x-ray PDC in 1T-TaS$_2$. A monochromatic x-ray pump beam is used to generate down-converted signal and idler photons that satisfy energy and momentum conservation through a reciprocal lattice vector G. A three-bounce Si(111) crystal analyzer, operated as a narrow-band energy filter, suppresses elastic scattering and isolates the weak PDC signal, which is recorded by a two-dimensional detector.

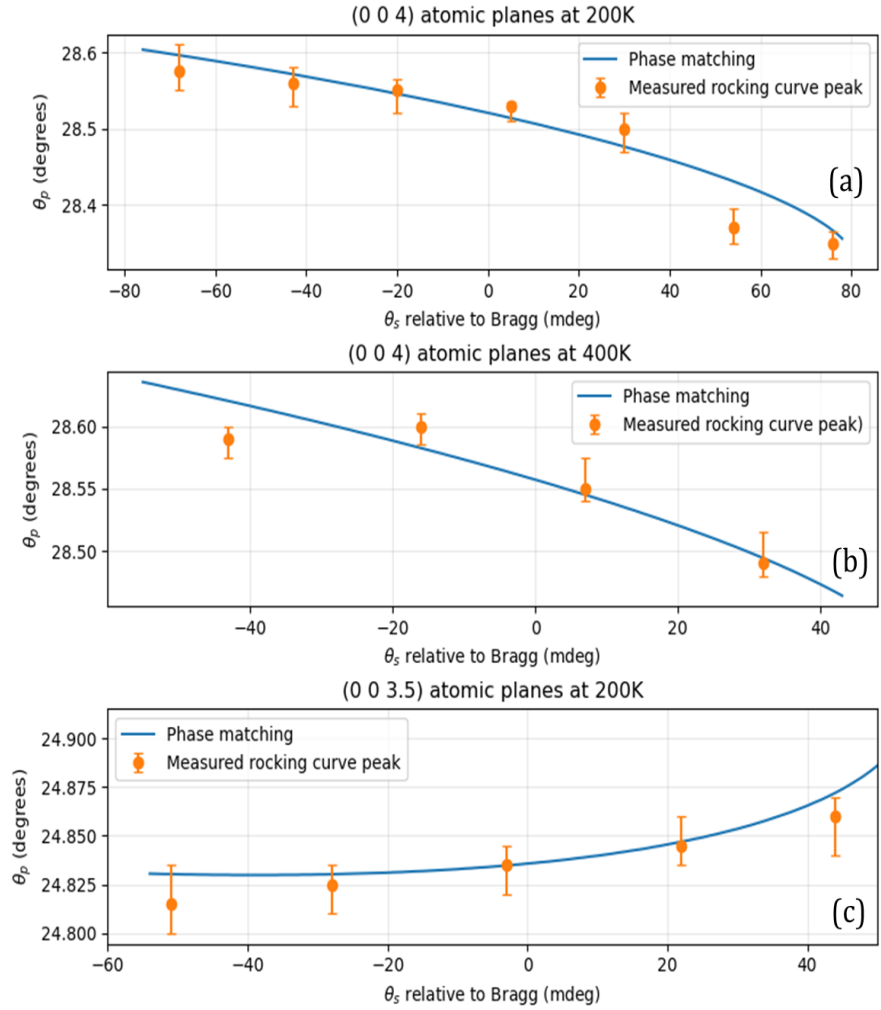

**Fig. 2**: Phase matching of x-ray PDC at an idler energy of 40 eV. Measured PDC rocking-curve peak angles versus detector angle for (a) the (0,0,4) reciprocal lattice vector at 200 K, (b) the (0,0,4) vector at 400 K, and (c) the (0,0,7/2) vector at 200 K. Solid lines show the calculated peak angles from the phase-matching condition. Error bars denote the fitted peak FWHM.

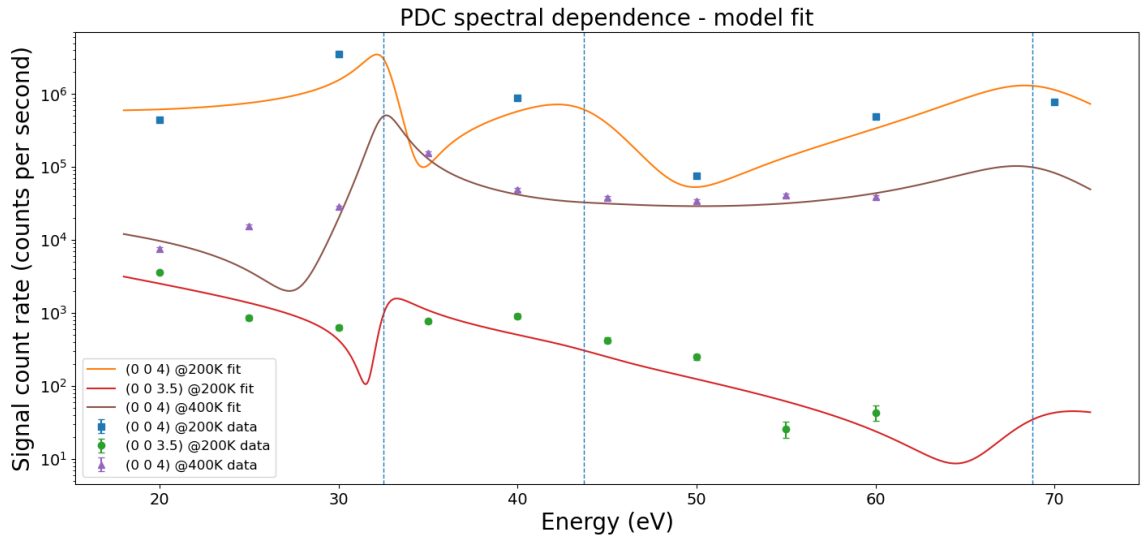

**Fig. 3**: Fits to the PDC spectra using a coherent Fano-type model for the nonlinear susceptibility. Data are shown for the (0,0,4) and (0,0,7/2) reflections in the NCCDW phase (200 K) and for the (0,0,4) reflection in the ICCDW phase (400 K). Dashed lines mark the Ta $O_3$, $O_2$, and $O_1$ resonances. The model captures asymmetric line shapes and reflection-dependent resonance strengths.

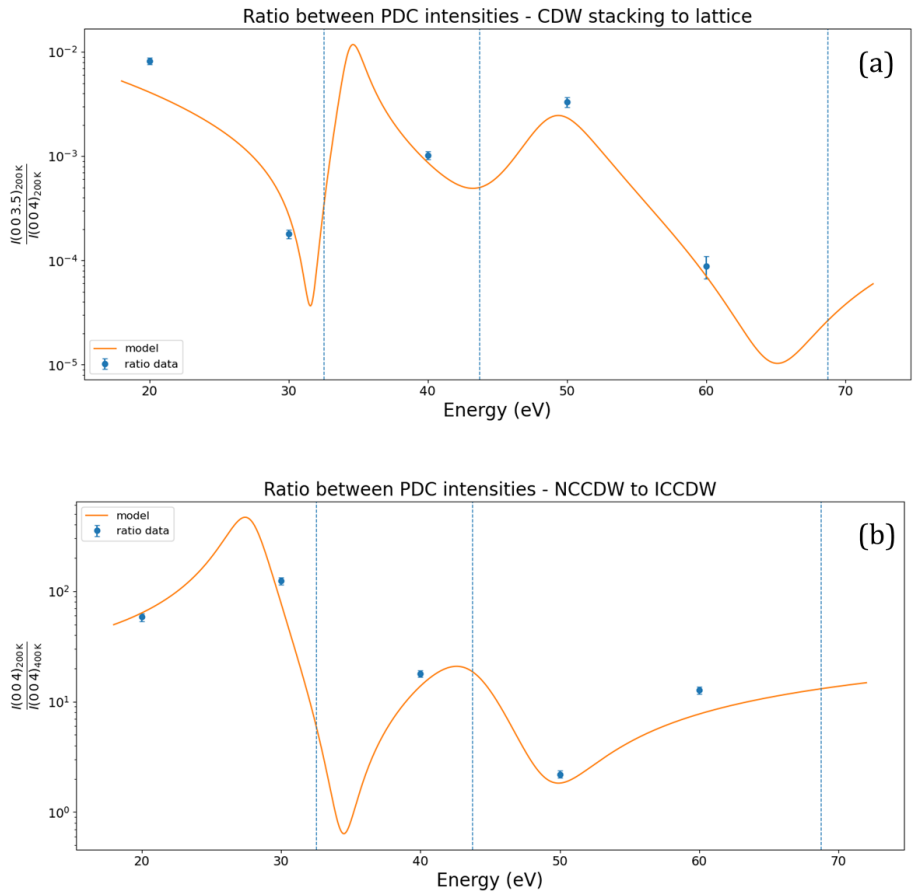

**Fig. 4**: Intensity ratios highlight phase and vector-dependent resonant PDC response. (a) Ratio of PDC intensity associated with the stacking-sensitive (0,0,7/2) reciprocal-lattice vector to that associated with the fundamental (0,0,4) vector at 200 K. (b) Ratio of PDC intensity associated with the (0,0,4) reciprocal lattice vector at 200 K to that at 400 K. Dashed lines indicate Ta O-shell resonances. Near resonance, the fundamental response is strongly enhanced while the stacking-sensitive signal remains suppressed, opposite to trends in elastic scattering.

# Supplementary Material for " Phase-Sensitive Nonlinear X-Ray Response in a Charge-Density-Wave Quantum Material"

**Authors:** S. Sofer[1], G. J. Man[2,3], A. Bombardi[4], and S. Shwartz[1]

**Affiliations:**

[1]*Physics Department and Institute of Nanotechnology, Bar-Ilan University, Ramat Gan, 52900 Israel*

[2]*Gabriel Man och Partners AB, Uppsala 752 57, Sweden*

[3]*Man Labs Technologies LLC, Dallas, Texas 75219, USA*

[4] *Diamond Light Source, Harwell Science and Innovation Campus, Didcot OX11 0DE, United Kingdom*

# Supplementary Note 1 – Rocking curves

We provide additional details on our data-analysis procedures and on the validation of the measured PDC signal. Representative rocking curves are shown whose peak positions agree with the calculated phase-matching angles, providing strong evidence that the detected emission originates from PDC. The angular dependence of the PDC efficiency is determined by the phase matching condition $\mathbf{k}_p + \mathbf{G} = \mathbf{k}_s + \mathbf{k}_i$. To compare experiment to theory we select a region of interest (ROI) on the detector centered at a fixed emission direction with an angular width of 0.01 degrees, thereby fixing $\mathbf{k}_s$ for a given reciprocal lattice vector $\mathbf{G}$. Using the measured photon energies and the known experimental geometry, we then solve the phase-matching relation to obtain the corresponding pump incidence (phase-matching) angle and compare it with the measured rocking-curve maximum.

Figures 1-3 show the measured rocking curves for the (0 0 4) reciprocal lattice vector at the NCCDW phase, for the (0 0 4) reciprocal lattice vector at the ICCDW phase, and for the (0 0 7/2) reciprocal lattice vector at the NCCDW phase, respectively. In all cases, the measured rocking-curve maxima agree with the predicted phase-matching angles within the ROI angular acceptance.

Next, we examine the energy dependence of the phase matching condition. For these measurements, the detector ROI has an angular with a width of 0.1 degrees. Supplementary figure 4 shows the rocking curves at idler energies 30,40, and 50 eV. In each panel, the shaded band denotes the calculated angles of the phase matching, with its width set by the ROI. The measured rocking-curve maxima again agree well with the calculated phase matching angles.

## Supplementary Note 2 – Analysis model

To describe the resonance-shaped dependence of the PDC intensities, we model each measured dataset $j$ (fixed reciprocal lattice vector and temperature) as a slowly varying envelope multiplying the square modulus of a coherent complex amplitude,

$$I_j(E) = C_j \, e^{-\alpha E} \, |A_j(E)|^2.$$

Here $E$ is the idler energy, $C_j$ is a reciprocal lattice vector and temperature dependent scale factor, and $\alpha$ describes a common, smooth falloff with energy that is not specific to a particular resonance. The resonant physics is contained in the complex amplitude $A_j(E)$, written as the coherent sum of a non-resonant background and three shallow core resonances,

$$A_j(E) = c_0 + \sum_{k=1}^{3} c_{j,k} \, F_k(E), \, F_k(E) = \frac{\varepsilon_k(E) + q}{\varepsilon_k(E) + i}, \, \varepsilon_k(E) = \frac{E - E_k}{\Gamma_k/2}.$$

The functions $F_k(E)$ are canonical Fano amplitudes, $E_k$ and $\Gamma_k$ are the resonance energies and effective widths (including instrumental broadening of 1 eV), $q$ is the Fano asymmetry parameter, and $c_{j,k}$ are real coefficients that quantify how strongly dataset $j$ couples to each resonant channel. In practice, $E_k$ and $\Gamma_k$ are fixed to values of the Ta core-level structure[1], and $q, c_0$ are fixed (to reduce parameter degeneracy with $c_{j,k}$). The fit then focuses on extracting the relative channel weights $c_{j,k}$ across reflections and temperatures. To emphasize physically meaningful differences between Fourier components and suppress common-mode factors, we fit the absolute intensities together with intensity ratios ($\frac{I_{003.5,200}}{I_{004,200}}$ and $\frac{I_{004,200}}{I_{004,400}}$). Supplementarty table 1 shows the parameters taken in the model, both fixed and fitted. Supplementary figs. 5-6 show the residuals defined as $r_I(E) = \frac{\log(I_{\text{measured}}(E)) - \log(I_{\text{model}}(E))}{\sigma_I}$ where $I_{measured}$ and $I_{model}$ are the measured and modelled intensities, respectively, and $\sigma_I$ is the estimated error at each measured intensity. for the fitted intensities and ratios on logarithmic scale. The model achieves a mean error of 0.2 decades (corresponding to an average factor of about $10^{0.2} \sim 1.6$) indicating that it captures the measured PDC intensities to within a constant-factor accuracy across the full energy range.

---

[1] M. F. Limonov, M. V. Rybin, A. N. Poddubny and Y. S. Kivshar, Nature Photonics **11**, 543-554 (2017).

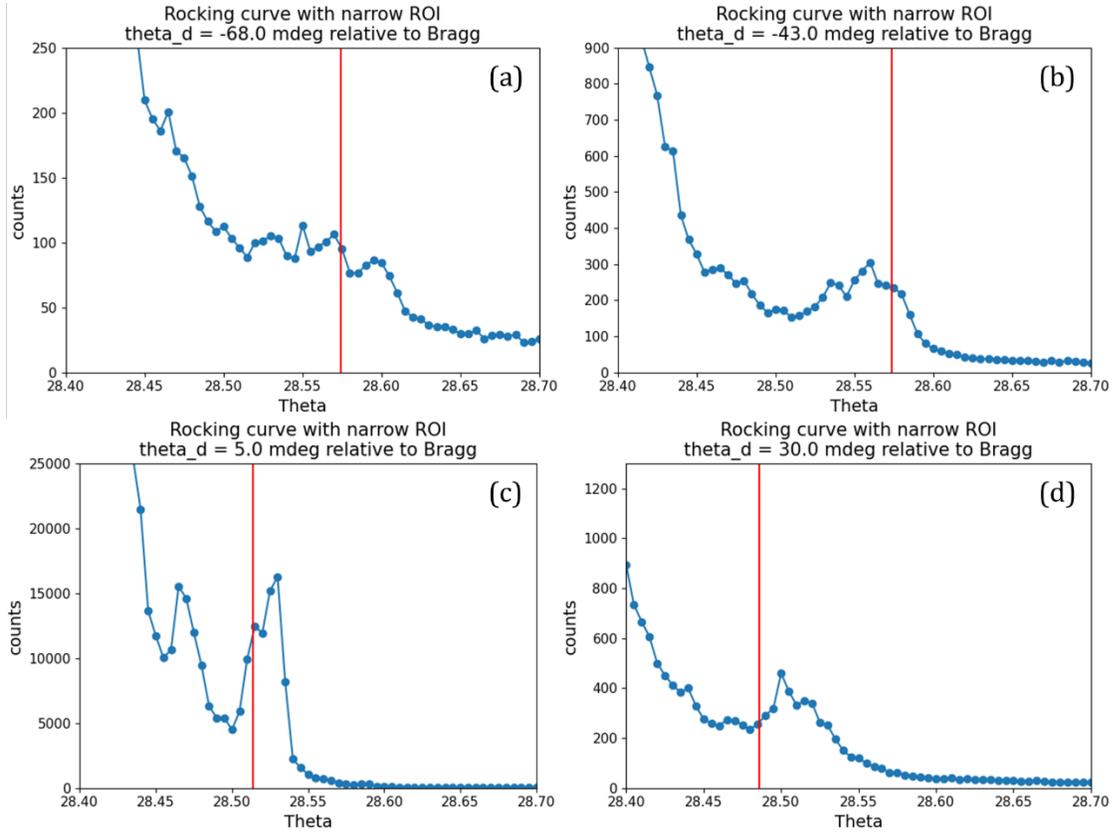

Supplementary Figure 1: **1T-TaS$_2$ rocking curve for the (0 0 4) reciprocal lattice vector at the NCCDW phase with an idler energy of 40 eV.** The tall peak at lower sample angle is the residual elastic scattering. The red line marks the calculated phase matching angle.

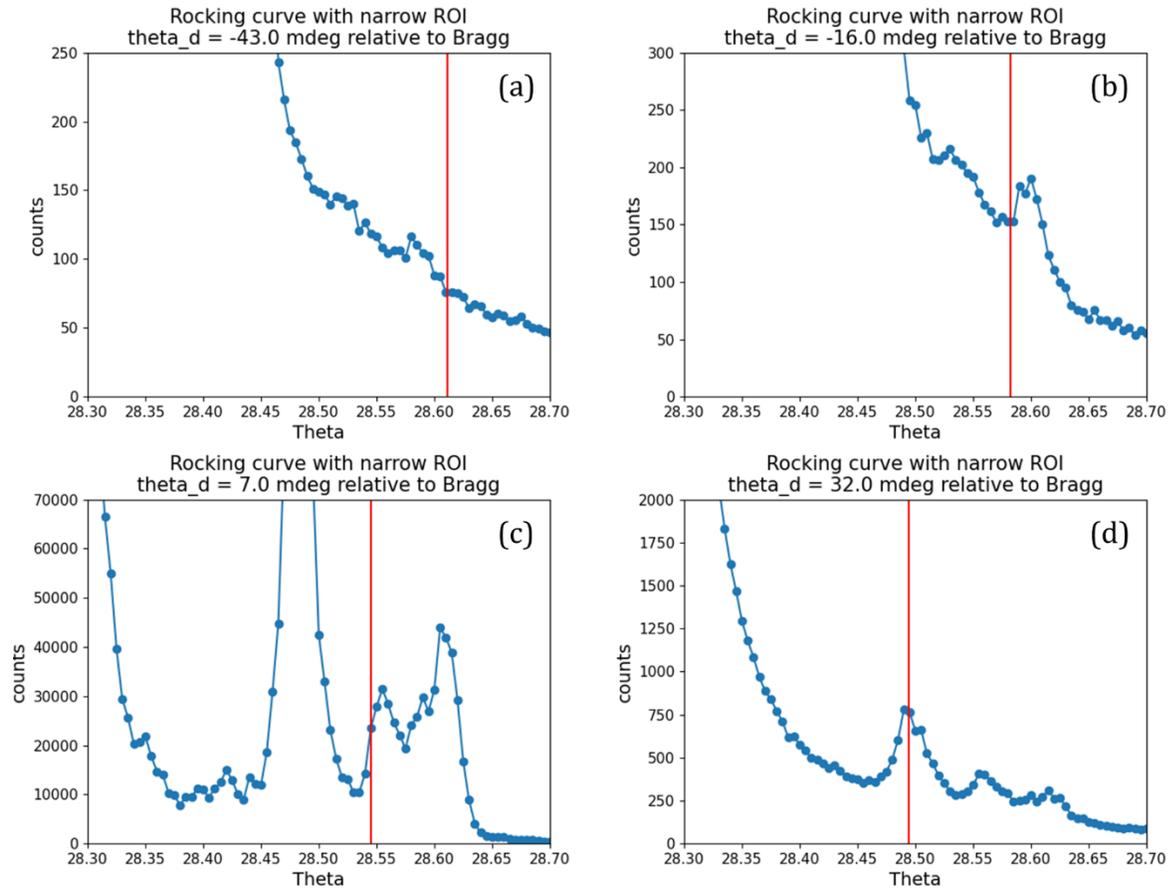

Supplementary Figure 2: **1T-TaS$_2$ rocking curve for the (0 0 4) reciprocal lattice vector the ICCDW phase with an idler energy of 40 eV.** The tall peak at lower sample angle is the residual elastic scattering. The red line marks the calculated phase matching angle.

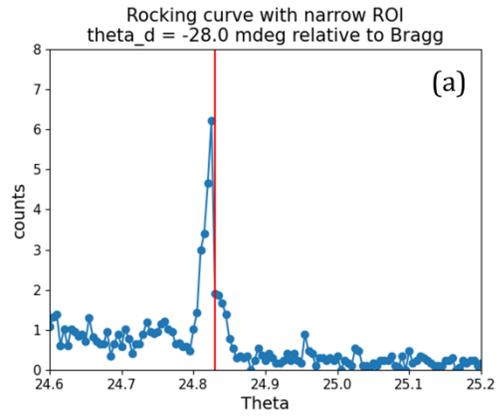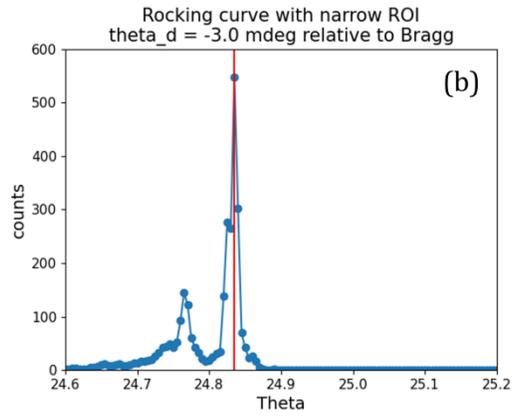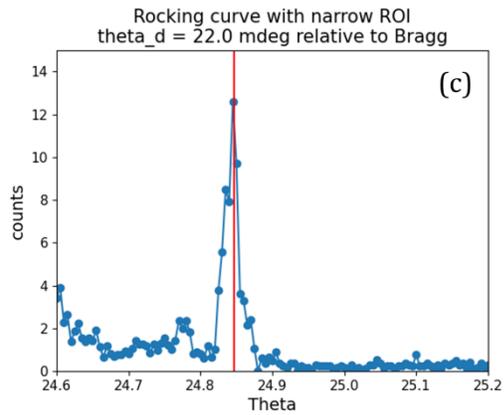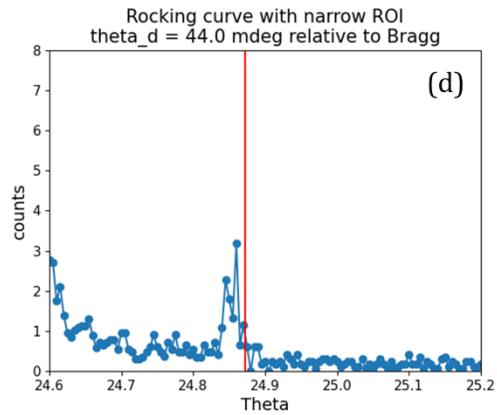

Supplementary Figure 3: **1T-TaS$_2$ rocking curve for the (0 0 7/2) reciprocal lattice vector at the NCCDW phase with an idler energy of 40 eV.** The tall peak at lower sample angle is the residual elastic scattering. The red line marks the calculated phase matching angle.

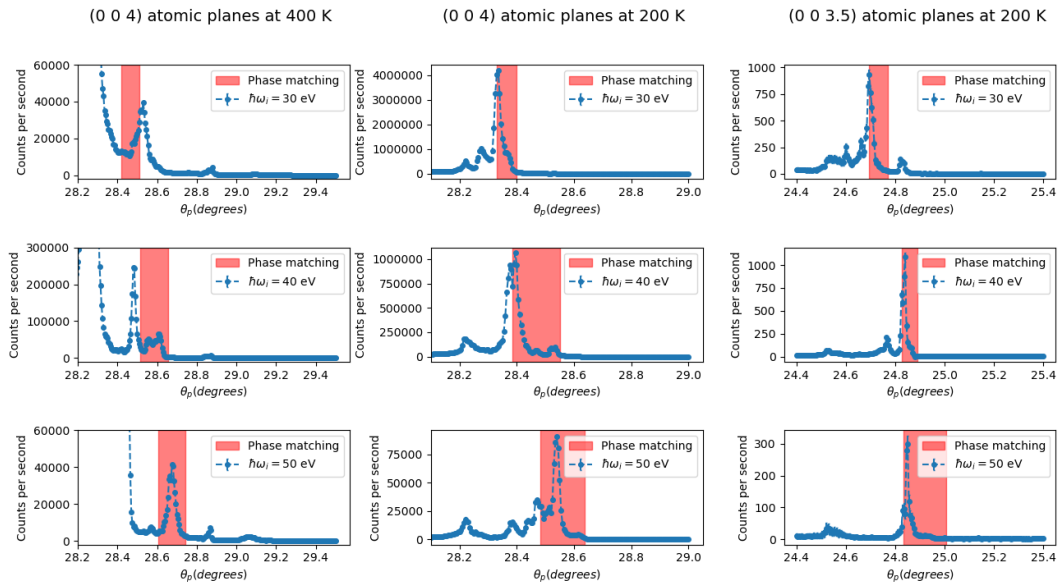

Supplementary Figure 4: **X-ray PDC rocking curves**. (a-c) Rocking curves at 400 K for the (0 0 4) vectors at idler energies of 30 eV, 40 eV and 50 eV, respectively. (d-f) Rocking curves at 200 K for the (0 0 4) vectors for idler energies of 30 eV, 40 eV and 50 eV, respectively. (g-i) Rocking curves at 200 K for the (0 0 3.5) vectors at idler energies of 30 eV, 40 eV and 50 eV, respectively. In each panel the red area indicates the predicted angle range with its width set by the ROI angular acceptance.

| parameter | value |
|---|---|
| a_004_200K | 15.7242 |
| a_0035_200K | 2.450121 |
| a_004_400K | 11.46022 |
| b_shared | -0.09637 |
| c1_004_200K | -0.19806 |
| c2_004_200K | -0.15744 |
| c3_004_200K | -0.80903 |
| c1_0035_200K | 2.899461 |
| c2_0035_200K | -0.19773 |
| c3_0035_200K | 2.85645 |
| c1_004_400K | 0.690558 |
| c2_004_400K | 0.010389 |
| c3_004_400K | -1.86573 |
| q_fixed | 2 |
| c0_fixed | 1 |
| E1_eV | 32.51 |
| E2_eV | 43.71 |
| E3_eV | 68.75 |
| G1_intr_eV | 1.6 |
| G2_intr_eV | 7.51 |
| G3_intr_eV | 8.88 |
| Gamma_beam_eV | 1 |
| G1_eff_eV | 1.886796 |
| G2_eff_eV | 7.576285 |
| G3_eff_eV | 8.936129 |
| f_sys | 0.05 |

| parameter | (0,0,4) @ 200K | (0,0,4) @ 400K | (0,0,7/2) @ 200K |
|---|---|---|---|
| C | 15.7242 | 11.46022 | 2.450121 |
| $\alpha$ | -0.096371462 | | |
| c1 | -0.19806 | 0.690558 | 2.899461 |
| c2 | -0.15744 | 0.010389 | -0.19773 |
| c3 | -0.80903 | -1.86573 | 2.85645 |
| q | 2 | | |
| c0 | 1 | | |
| E1 (eV) | 32.51 | | |
| E2 (eV) | 43.71 | | |
| E3 (eV) | 68.75 | | |
| G1_eff_eV | 1.886796226 | | |

| | |
|---|---|
| **G2_eff_eV** | 7.576285369 |
| **G3_eff_eV** | 8.936128916 |

**Supplementary Table 1**: Fitted and fixed parameters used in the coherent Fano-interference model. Grey-shaded entries denote parameters optimized in the fit, while unshaded entries were held fixed.

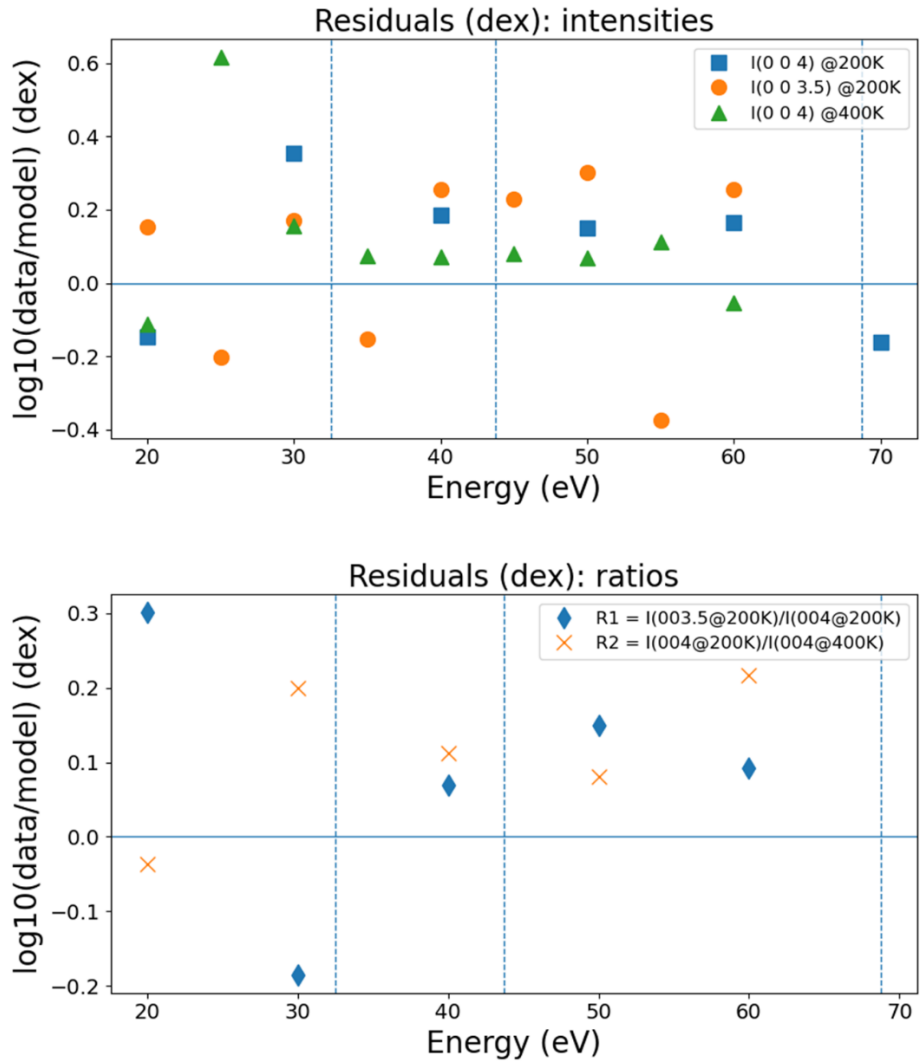

Supplementary Figure 5: **Logarithmic residuals of the coherent Fano-interference model. Residuals are plotted as $\log_{10}(I_{\text{data}}/I_{\text{model}})$ in units of decades (dex).** Positive values indicate the model underestimates the data. (a) Residuals for the absolute PDC intensities: (0,0,4) at 200K, (0,0,7/2) at 200 K, and (0,0,4) at 400 K. (b) Residuals for the intensity ratios evaluated at common energies: $R_1 = I_{(0,0,7/2),200K}/I_{(0,0,4),200K}$ and $R_2 = I_{(0,0,4),200K}/I_{(0,0,4),400K}$. Vertical dashed lines mark the resonance energies $E_1$, $E_2$, and $E_3$ used in the model.